\documentclass[a4paper,fleqn]{cas-sc}

\usepackage[authoryear,longnamesfirst]{natbib}
\usepackage{xcolor} 
\usepackage{subcaption}
\usepackage{bm}
\usepackage{multirow}
\usepackage{booktabs}
\usepackage{float}

\usepackage{amssymb}
\usepackage{soul}
\usepackage[normalem]{ulem}
\usepackage{amsmath}

\usepackage{lineno}

\begin{document}
\let\WriteBookmarks\relax
\def\floatpagepagefraction{1}
\def\textpagefraction{.001}

\shorttitle{Plutino and KBO forced planes}    

\shortauthors{Matheson, Malhotra}  

\title [mode = title]{The forced orbital planes of the Plutinos and other resonant KBOs}  



%

\author[1]{Ian C. Matheson}[orcid=0000-0003-0940-7176]

\cormark[1]


\ead{ianmatheson@arizona.edu}


\credit{Investigation, Formal analysis, Software, Visualization, Data curation, Writing - original draft, Writing - review \& editing}

\affiliation[1]{organization={Department of Aerospace and Mechanical Engineering, The University of Arizona},
            addressline={1130 N. Mountain Ave., P.O. Box 210119}, 
            city={Tucson},
            postcode={85721}, 
            state={AZ},
            country={USA}}

\author[2]{Renu Malhotra}
[orcid=0000-0002-1226-3305]



\credit{Conceptualization, Funding acquisition, Supervision, Writing - review \& editing}

\affiliation[inst2]{organization={Lunar and Planetary Laboratory, The University of Arizona},
            addressline={1629 E. University Blvd.}, 
            city={Tucson},
            postcode={85721}, 
            state={AZ},
            country={USA}}


\cortext[cor1]{Corresponding author}


\begin{abstract}
The forced plane of non-resonant minor planets has been understood to be 
well-estimated by the local Laplace plane given by linear Laplace-Lagrange secular perturbation theory, but there is no extant theory for the forced plane of resonant minor planets. 
With improved methodology, we revisit our previous measurement of the mean orbital plane of the observed Plutinos, a group of Kuiper belt objects locked in Neptune’s 3:2 mean motion resonance. 
In the J2000 ecliptic-equinox reference frame, the measured mean plane has an inclination of $1.9^\circ$ and an ascending node at $46.8^\circ$ longitude, with a 95\% confidence of $2^\circ$ in the plane position; it is well separated from the local Laplace plane but \textcolor{blue}{is} indistinguishable from the solar system's invariable plane. 
We also show that, at high statistical confidence, the two dynamical subgroups, the doubly resonant Plutinos whose arguments of perihelion librate about $\pm90^\circ$, have different mean planes from each other.
Additionally, we measure the mean planes of four other observed resonant groups in the Kuiper belt: the 5:3, 7:4, 2:1, and 5:2. 
Although these groups have substantial sample sizes (in the range 56--105), their mean planes are presently not statistically distinguishable from their local Laplace plane, Neptune's plane and the invariable plane. 
For all five resonant groups, we also report the best-fit von Mises function for their inclinations.
These empirical measurements underscore the need for theoretical analysis to understand the spatial dynamics of resonant populations of minor planets.
Such analysis could potentially enable their use for uncovering unmodeled perturbations and/or undiscovered perturbers.
\end{abstract}


\begin{highlights}
\item We estimate the forced orbit planes of the Plutinos and other resonant Kuiper belt populations from the debiased mean planes of the observed populations.
\item The Plutinos mean plane is distinct from the Laplace plane (as estimated by linear Laplace-Lagrange secular theory), but the others are not.
\item The mean planes of  resonant KBOs outside the 3:2 MMR are similar to those of nearby non-resonant populations.
\end{highlights}

\begin{keywords}
 Plutinos \sep Orbital resonances \sep Kuiper belt dynamics
\end{keywords}

\maketitle


\section{Introduction}
\label{s:introduction}
Gravitational perturbations by massive planets are well understood to shape the long-term orbital evolution of small bodies in the Solar System.
Over time, exchanges of energy and angular momentum change the shapes, sizes, and orientations of small body orbits and of planetary orbits as well.
In particular, the time variation of the orientation of the orbital plane of a small Solar System body (such as an asteroid or a Kuiper Belt object) is described by the Laplace-Lagrange linear secular theory \citep{md99}.
In this theory, the orientation of a small body's orbital plane under the influence of $N$ massive planets is described with a superposition of a ``free inclination'' mode and $N$ secular inclination modes, one of which has zero frequency and is associated with the invariable plane.
From this description, we can infer that at any given epoch, a group of small bodies within a narrow range of orbital semimajor axis will have randomly oriented ``free inclinations'' superposed on a forced plane determined by the planets' modes.
This forced plane is called the ``Laplace plane'', and it is expected to coincide, to within statistical variation, with the mean plane of small bodies in a narrow range of semi-major axis \citep{cc08}.

This theoretical prediction has been successfully tested with observational data of non-resonant populations of asteroids \citep{Cambioni:2018} and Kuiper belt objects \citep[e.g.,][]{Brown:2004b,Chiang:2008,vm17,mm23}.
However, Laplace-Lagrange linear secular theory breaks down inside mean motion resonances, and there is no alternative
analytical theory to describe the forced plane of resonant populations.
Owing to this gap in dynamical theory, a number of recent studies 
that seek signatures of undiscovered distant planets in the mean planes of small body populations either restrict their data analysis to only non-resonant objects \citep[e.g.,][]{ossos14,vm17,siraj2025measuring}, or to the small numbers of objects deemed to be too distant to be affected by weak mean motion resonances in the distant trans-Neptunian region \citep[e.g][]{pichierri2025measuring,batygin2019planet,brown2021orbit}.

We are attempting to fill this gap in our understanding of the mean and forced planes of resonant populations with empirical determinations of the mean planes of some of the largest cataloged small body populations in planetary mean motion resonances.
In a previous paper \citep{mm26hildas}, we investigated this question by examining the Hilda asteroids in Jupiter's interior 3:2 MMR.
We chose the Hildas because they are the largest observationally complete resonant population in the Solar System (3893 objects with $H\leq16.3$), so their mean plane and its statistical confidence can be calculated with straightforward orbit-normal averaging and von Mises-Fisher (vMF) statistics, without need of special procedures to account for survey biases in the observed population.
For this sample, we found that the mean plane could be measured (at 95\% confidence) with a precision of $0.2^\circ$. 
We also found that the 95\% confidence region for the mean plane encompasses the orbital plane of Jupiter and the local Laplace plane predicted from linear secular theory, but does not enclose the invariable plane of the Solar System.
This rules out the invariable plane as a viable candidate for the Hildas’ forced plane at the current epoch, but it does not rule out Jupiter’s plane or the local Laplace plane defined by the four giant planets. 
With N-body numerical simulations, we confirmed that over time the Hilda group’s mean plane continues to remain distinct from the invariable plane, and that it continues to remain close to the Jupiter plane and slightly closer to the Laplace plane. 
In the case of the Hildas, the close proximity of the local Laplace plane to Jupiter’s plane requires a much higher precision of mean plane measurement to determine which of these two planes might be the true forced plane of the resonant asteroids.

In this paper, we examine the Plutinos, a dynamical group in Neptune's exterior 3:2 MMR.
The Plutinos are the largest resonant group in Neptune's exterior MMRs, and are of great interest for their utility in testing models of Solar System formation.
Because current catalogs of such distant objects are subject to strong observational biases, we use a statistical method recently described by \citet{siraj2025measuring} to compute their mean plane and its statistical uncertainty.
This method is nearly insensitive to survey biases and does not require the construction of a synthetic or model population.
We compare the mean plane calculated using this method with our previous estimate of the Plutino mean plane \citep{mmk23} and compare both mean plane estimates with a third method from \citet{vm17}.
A substantial fraction of the Plutinos are also subject to the von Zeipel-Lidov-Kozai resonance (vZLK), which results in coupled oscillations in eccentricity and inclination alongside libration in argument of perihelion, $g$; there are two libration centers, so we compute the mean planes of these two subgroups of Plutinos separately. We also report the distribution of inclinations relative to the computed mean planes.
We also use the method of \citet{siraj2025measuring} to compute the mean planes of the next four most heavily populated Neptunian mean motion resonances, which in order of increasing semimajor axis are the 5:3, 7:4, 2:1, and 5:2 MMRs.
Because these MMRs do not have substantial vZLK populations, we do not calculate the mean planes of their few vZLK librators separately.

We present the results of our investigation by organizing the rest of this paper as follows.
\begin{itemize}
\item In section \ref{s:mean-plane-statistics}, we summarize three methods for mean plane estimation and we compare them to the statistics used in our previous look at the Plutinos \citep{mmk23}.
\item In section \ref{s:Plutinos}, we describe the cataloged Plutino sample and its mean plane. We discuss comparisons with three reference planes and  discuss the inclination distribution relative to the measured mean plane.
\item In section \ref{sec:appendix_mmrs}, we present the mean planes of the 5:3, 7:4, 2:1, and 5:2 MMRs and discuss their locations relative to the same three reference planes.
\item Section \ref{s:summary} provides a summary of the results and discusses prospects for improved accuracy and precision of mean plane measurements with forthcoming observational surveys.
\item In Appendix \ref{sec:appendix_lawler}, we discuss the mean plane of a synthetic Plutino population as a point of comparison to the cataloged population.
\end{itemize}

The three reference planes mentioned in Sections \ref{s:Plutinos}--\ref{sec:appendix_mmrs} are Neptune's orbital plane, the invariable plane of the Solar System, and the linear Laplace-Lagrange secular theory estimate of the local forced plane; we refer to the latter as the Laplace plane.

\section{Three ways to calculate the mean plane}
\label{s:mean-plane-statistics}

If a population is observationally complete and their unit vectors normal to the orbit plane, $\hat{\mathbf{J}}_i$, pass statistical tests for 
azimuthal symmetry about the mean direction
on the unit sphere, 
then the unbiased mean plane and its statistical uncertainty can be calculated using von Mises-Fisher (vMF) statistics, as summarized in \citet{mmk23}.
In that case, the mean plane is the simple sum of the individual orbit normal vectors, normalized to the surface of the unit sphere.
We call this the vMF method.
This produces a concentration parameter $\gamma_{\rm{est}}$.
The vMF distribution is locally similar to a bivariate normal distribution of standard deviation $\sigma=\gamma^{-0.5}$; it is also locally equivalent to a Rayleigh distribution of inclinations with scale parameter $\sigma=\arcsin\gamma^{-0.5}$.
The contours of the sampling distribution of the mean plane take the form of small circles centered on the mean plane.
The 95\% confidence region for the mean plane is a small circle with angular radius $\theta_{95\%}$ that contains 95\% of the probability mass for the sampling distribution of the mean plane.
As mentioned in section 1, we previously adopted this method for measuring the mean plane of the observationally complete population of the Hilda asteroids.

Because current catalogs of Kuiper belt populations are neither observationally complete nor unbiased, some care is needed to compute useful statistics.
To reduce the effects of observational biases on population-level analyses, some studies restrict the observational dataset to that produced by a single well-characterized survey, such as the Outer Solar System Origins Survey (OSSOS) \citep{ossos1}.
Single well-characterized surveys typically only detect and track a small fraction of the overall observed population, so their increased statistical power is limited by a smaller sample size.

Fortunately, a method exists to measure the mean orbital plane of a population of small bodies that is relatively insensitive to survey biases.
This method allows the mean plane of the Plutinos to be calculated from the entire observed population without detailed analysis of the observational history of the Plutino catalog.
Happily, it also allows the confidence in the mean plane to be calculated with only minimal assumptions about the source population.
First described by \citet{siraj2025measuring}, the calculation can be summarized as follows.
We call this the ``\textsc{sct}25 method.''

Begin with an observed population of $N$ small bodies, drawn from a much larger source population.
\citet{siraj2025measuring} make the ansatz that the inclination distribution of the source population is separable from that of the joint distribution of the orbital energy and total angular momentum.
They further assume that the orbital poles are azimuthally symmetric about a mean pole, and are statistically described by the von Mises-Fisher (vMF) distribution on the surface of the unit sphere \citep{fisher_1993}.

It should be understood that these assumptions apply to the notional \textit{source population}.
The purpose of this method is to calculate the \textit{debiased} mean plane of a heavily biased \textit{observed population}, so it is not to be expected that the \textit{observed population} will satisfy the conditions of azimuthal symmetry about the mean plane, and independence of inclination with respect to energy and scalar angular momentum.
An \textit{unbiased} sample from the source population can be expected to satisfy those conditions, but the mean plane of an unbiased sample is much more readily calculated with vMF statistics.

With these assumptions in mind, let the source population have a mean plane with unit normal vector $\hat{\mathbf{m}}$ and a vMF distribution with concentration parameter $\gamma$.
The unit orbital angular momentum vector (or the unit vector normal to the orbital plane) of the $i$th object is found from the position vector $\mathbf{r}_i$ and velocity vector $\mathbf{v}_i$ as $\hat{\mathbf{J}}_i=(\mathbf{r}_i\times\mathbf{v}_i)/||\mathbf{r}_i\times\mathbf{v}_i||$.
The angle $\theta_{i}$ between the $i$th object's position vector $\mathbf{r}_i$ and the mean plane unit normal vector is given by $\cos\theta_{i}=\hat{\mathbf{m}}\cdot\hat{\mathbf{r}}_i$, where $\hat{\mathbf{r}}_i=\mathbf{r}_i/||\mathbf{r}_i||$.
The likelihood $L$ of obtaining the observed sample of unit orbital angular momentum vectors $\mathbf{J}_i$, $i=1,...,N$, is described by
\begin{equation}
\log L=\gamma\,\hat{\mathbf{m}}\cdot\sum_{i=1}^N \hat{\mathbf{J}}_i-\sum_{i=1}^N \log I_0 \left(\gamma \sin\theta_{i}\right),
\label{e:logL}
\end{equation}
where $I_0 (\cdot)$ is the modified Bessel function of the first kind of order zero.

To estimate the mean plane of the population from the observed sample, one presupposes that the likelihood will be greatest when reckoned with respect to the true mean plane of the population.
Therefore, one varies $\hat{\mathbf{m}}$ and $\gamma$ in Eq. \ref{e:logL} to maximize the result.
Then $\hat{\mathbf{m}}_{\rm{est}}$ and $\gamma_{\rm{est}}$ denote the mean plane and the vMF concentration parameter that produce the maximum log-likelihood value, $\log L_{\rm{est}}$.

The statistical uncertainty in the
measurement of the mean plane is then found as follows.
If the true population is hypothesized to have mean plane $\hat{\mathbf{m}}_{\rm{ref}}$ and concentration parameter $\gamma_{\rm{est}}$, then
the measurement's significance relative to the reference plane $\hat{\mathbf{m}}_{\rm{ref}}$, written in terms of standard deviations, is
\begin{equation}
S_{\rm{ref,est}}=\sqrt{2(\log L_{\rm{est}}-\log L_{\rm{ref}})},
\end{equation}
where $\log L_{\rm{ref}}$ is calculated from Eq. \ref{e:logL} with $\hat{\bm{m}}=\hat{\mathbf{m}}_{\rm{ref}}$ and $\gamma=\gamma_{\rm{est}}$.
The significance $S_{\rm{ref,est}}$ has an asymptotic $\chi^2$ distribution with two degrees of freedom, giving a ``false alarm probability'' \citep{siraj2025measuring} of
\begin{equation}
P_{\rm{ref,est}}=\exp\left(-\frac{1}{2}S^2\right),
\end{equation}
which simplifies to
\begin{equation}
\label{e:pval}
P_{\rm{ref,est}}=\frac{L_{\rm{ref}}}{L_{\rm{est}}}.
\end{equation}
The false-alarm probability $P_{\rm{ref,est}}$ is interpreted as
the probability of observing a sample of $N$ objects with mean plane $\hat{\mathbf{m}}_{\rm{est}}$ and concentration parameter $\gamma_{\rm{est}}$
, given the assumed mean plane $\hat{\mathbf{m}}_{\rm{ref}}$ and conditioned upon the concentration parameter $\gamma_{\rm{est}}$.
Because $L_{\rm{est}}$ is found by maximizing Eq. \ref{e:logL}, $L_{\rm{ref}}<L_{\rm{est}}$, so $0<P_{\rm{ref,est}}<1$ for all $\hat{\bm{m}}_{\rm{ref}}$.
If $P_{\rm{ref,est}}$ is greater than some threshold value, say $P\geq0.05$, then we say that it is statistically credible at the 5\% level that the true mean plane of the source population is the hypothesized reference plane $\hat{\mathbf{m}}_{\rm{ref}}$.
To construct a credible region for the sample mean plane $\hat{\mathbf{m}}_{\rm{est}}$, we vary the first two elements of $\hat{\mathbf{m}}$ (i.e., $q$ and $p$ as in Figure \ref{fig:compare_methods_Plutinos}) over a grid centered on $\hat{\mathbf{m}}_{\rm{est}}$ while holding $\gamma$ constant at $\gamma=\gamma_{\rm{est}}$, and we evaluate Eq. \ref{e:pval} on that grid.
Following \citet{siraj2025measuring}, we define the 95\% credible region as the contour ellipse where $P_{\rm{ref,est}}=0.05$, and we define its mean angular width $\theta_{95\%}$ as the geometric mean of the semimajor and semiminor axes of that ellipse.
We use the language of ``credibility'' rather than ``confidence'' to express the uncertainty in the mean plane because the \textsc{sct25} method is a Bayesian approach based on a vMF prior distribution.
Note that this 95\% credible region is actually defined for a \textit{conditional} posterior distribution with $\gamma=\gamma_{\rm{est}}$ rather than for a \textit{marginal} posterior distribution obtained by integrating the probability distribution corresponding to Eq. \ref{e:logL} over all possible values of $\gamma$.
The use of a \textit{conditional} rather than \textit{marginal} false-alarm probability and credible region is a limitation of the \textsc{sct25} method; we found that marginalizing over $\gamma$ by computing its optimal value in Eq. \ref{e:logL} for every point $\hat{\bm{m}}$ in a grid search is computationally prohibitive.

The \textsc{sct25} method's insensitivity to sample bias lies in writing the probability of observing the $i$th object as a product of three factors: 
a pointing function $w(\mathbf{r}_i)$ that contains all biases introduced by the position $\mathbf{r}_i$ of the object at the time of the observation, a spherically symmetric factor $g(E_i,J_i)$ that accounts for the size and shape of the orbit via its energy $E_i$ and scalar angular momentum magnitude $J_i$, and a rotationally symmetric factor $f(\hat{\bm{J}}_i\cdot\hat{\bm{m}}_i)$ that accounts for the object's inclination relative to the mean plane.
Note that $\hat{\bm{J}}_i\cdot\hat{\bm{m}}_i=\cos{i_{\rm{rel},i}}$.
When this probability of observation is transformed into a conditional probability and then a likelihood by normalizing by the integrated probability of observations over the range of possible relative inclinations, the pointing function $w$ and energy function $g$ are factored out of the integral and thus canceled from the probability of observation (in the numerator) and the normalizing factor (in the denominator).
This leaves only the (unbiased) relative inclination factor $f$.

An alternative method for computing the debiased mean plane of a biased sample relies on finding the plane of symmetry of the sky-plane velocity vectors of the observed objects, as described by \citet{bp04}, then following the procedure in \citet{vm17} to construct synthetic samples from small perturbations of the real observed objects to build a sampling distribution and confidence region for the mean plane.
We call this the \textsc{vm17} method.
We report the mean angular width $\theta_{95\%}$ of the 95\% confidence ellipse for the \textsc{vm17} method as the geometric mean of the ellipse semimajor and semiminor axes.

When summarizing the \textsc{sct25} mean plane computation, we spoke of the unit orbital angular momentum vector $\hat{\bm{J}}$. This is normal to the orbital plane, and is written as
\begin{equation}
\hat{\bm{J}}=(J_x,J_y,J_z)=(\sin{i}\sin{\Omega},-\sin{i}\cos{\Omega},\cos{i}),
\label{e:Jhat_definition}
\end{equation}
where the inclination $i$ and longitude of ascending node $\Omega$ will be measured in the J2000 reference frame.
However, in the remainder of this paper we prefer to adapt notation from the presentation of linear secular theory in \citet{md99} and write the location of the mean plane as
\begin{equation}
\hat{\bm{q}}=(q,p,s)=(-J_y,J_x,J_z)=(\sin{i}\cos{\Omega},\sin{i}\sin{\Omega},\cos{i}).
\label{e:qps_definition}
\end{equation}

We report and plot the mean plane components $(q,p)$ rather than as $(J_x,J_y)$ because it is more intuitive to read $\Omega$ directly from a plot of $(q,p)$ than from a plot of $(J_x,J_y)$.

\citet{siraj2025measuring} used Monte Carlo simulation to show that their method, when applied to biased Kuiper belt surveys, typically returns a slightly more accurate mean plane than the \textsc{vm17} method and a dramatically more accurate mean plane than the vMF method.
In this paragraph, we summarize the strengths and weaknesses of these three mean plane calculations.
The vMF mean plane and confidence region are computationally inexpensive to calculate even with very large sample sizes; they are reliable for observationally complete samples that are rotationally symmetric about their mean plane, but unreliable for biased samples.
The \textsc{vm17} mean plane can be calculated fairly quickly as a point estimate, but the confidence region requires efficient numerical implementation and demands a large amount of computation with multiple assumptions about the source population.
It is cumbersome, but scales well to large samples.
The maximum-likelihood \textsc{sct25} mean plane is insensitive to sample bias, but it requires a black-box optimization step that 
(in our Python implementation using \texttt{scipy.optimize.minimize})
can be heavily demanding in computational resources for sample sizes larger than a few thousand objects.
After the mean plane is found, its credible region can be computed using a relatively quick, deterministic, grid search method without further use of numerical optimization.
Both the vMF method and the \textsc{sct25} method return an estimate of the concentration parameter $\gamma_{\rm{est}}$ of the  vMF distribution of the inclinations relative to the best-fit mean plane as part of the mean plane computations; the \textsc{vm17} method does not, as it seeks the plane-of-symmetry of the sky-plane velocity vectors, without assuming a vMF distribution about that symmetry plane.

\subsection{Three mean planes of the cataloged Plutinos}
\label{ssec:three_mean_planes}

\begin{figure*}
\includegraphics[width=3in]%
{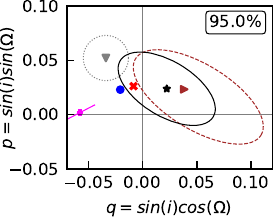}
\caption{
Mean plane estimates for the observed Plutinos: 
the black $\bigstar$ indicates the debiased estimate using \textsc{sct25},
the brown {\tiny $\blacktriangleright$} indicates the debiased estimate using \textsc{vm17};
95\% confidence/credibility regions are also indicated for each of these.
For comparisons, we also indicate the following planes:
the vMF mean plane (gray $\blacktriangledown$),
the invariable plane (red $\times$),
Neptune's plane (blue $\bullet$),
the Laplace plane (magenta $+$) and its range over the Plutinos' range of semimajor axis (magenta line segment).
The reference frame is the J2000 ecliptic-equinox.
}
\label{fig:compare_methods_Plutinos}
\end{figure*}

\begin{table}
    \begin{tabular}{lrrrrlll}
    \hline\hline  \addlinespace[2pt]
    \noalign{\hglue-4truein{Mean plane locations}} 
    \addlinespace[1pt]
    & $q$ & $p$ & $i$ & $\Omega$ & $\theta_{95\%}$ & $\gamma_{\rm{est}}$ & $\sigma_{\rm{est}}$ \\ 
    \addlinespace[1pt]
    vMF & -0.034 & 0.052 & $3.57^\circ$ & $122.93^\circ$ & $1.19^\circ$ & 31.5 & $10.2^\circ$\\
    \textsc{sct25} & 0.022 & 0.024 & $1.88^\circ$ & $46.83^\circ$ & $2.04^\circ$ & 26.4 & $11.2^\circ$ \\
    \textsc{vm17} & 0.038 & 0.023 & $2.58^\circ$ & $31.30^\circ$ & $2.59^\circ$ & $-$ & $-$ \\ 
    \addlinespace[3pt]
    \hline \addlinespace[3pt]
    \noalign{\hglue-3.2truein{Mean plane separations (in degrees)}} 
    \addlinespace[3pt]
          & vMF  & \textsc{sct25} & \textsc{vm17} & Invariable & Neptune & Laplace \\ \addlinespace[1pt]
                    vMF & 0    & 3.61 & 4.46 & 2.09 & 1.84 & 3.19 \\
        \textsc{sct25} & 3.61 & 0    & 0.92 & 1.77 & 2.46 & 4.76 \\
        \textsc{vm17}  & 4.46 & 0.92 & 0    & 2.69 & 3.38 & 5.65 \\ 
        \addlinespace[3pt]
        \hline \addlinespace[3pt]
         \noalign{\hglue-4.6truein{$P$-values}} 
         \addlinespace[3pt]
            &  &  &  & Invariable & Neptune & Laplace \\ \addlinespace[1pt]
        vMF &  &  &  & 0.00010 & 0.00081 & 0 \\
        \textsc{sct25} &  &  &  & 0.17 & 0.018 & 0 \\
        \textsc{vm17} &  &  &  & 0.024 & 0.0023 & $1.1\cdot10^{-9}$ \\ \addlinespace[3pt]
        \hline
    \end{tabular}
    \caption{
Data accompanying Figure \ref{fig:compare_methods_Plutinos}.
The first block reports the locations of the observed Plutinos’ mean plane calculated with three different methods, in the J2000 reference plane;
the last columns tabulate the concentration parameter of the best-fit vMF function for the inclinations relative to the mean plane and the corresponding width of the Rayleigh relative inclination distribution.
The second block reports the angular separations amongst the three planes, as well as their separations from three physically relevant planes (the invariable plane, Neptune’s plane and the Laplace plane).
The third block reports the $P$-values of the angular separations from the fixed reference planes.
See main text for details.
}
\label{t:figure1_meanplanes}
\end{table}

We demonstrate the differences in performance between these three methods by applying them to the observed Plutino population described in section \ref{s:Plutinos}.
The observed Plutino population does not satisfy the conditions of azimuthal symmetry about the mean plane and independence of inclination with respect to orbital energy and angular momentum magnitude, but those are qualities ascribed to the \textit{source} population by the \textsc{sct25} method, not to the \textit{observed sample}.
In Figure \ref{fig:compare_methods_Plutinos}, the \textsc{sct25} mean plane (black $\bigstar$) and \textsc{vm17} mean plane (brown {\tiny $\blacktriangleright$}) closely agree on the location of the Plutino mean plane, but the vMF estimate (gray $\blacktriangledown$) is distant from either.
The mean plane locations in Figure \ref{fig:compare_methods_Plutinos} are also shown in Table \ref{t:figure1_meanplanes}.
This table also contains the angular distances between them and the $P$-values of various reference planes relative to the mean plane estimates.
These $P$-values represent the sample mean probability distribution's mass outside an ellipse passing through the reference point.
The rows are the different mean plane probability distributions, and the columns are the reference planes.
For example, according to \textsc{sct25} statistics, 
17\% of the probability mass for the probability distribution of the mean plane of the Plutinos lies outside an ellipse centered on the \textsc{sct25} mean plane and passing through the invariable plane, and 1.8\% lies outside an ellipse passing through the orbital plane of Neptune.
Therefore, the \textsc{sct25} mean plane is not statistically distinct from the invariable plane at the 5\% level (95\% credibility), but it is statistically distinct from the plane of Neptune at better than the 5\% level (95\% credibility).

According to vMF statistics (Table \ref{t:figure1_meanplanes}, third block, first row), the mean plane of the observed Plutinos (black $\bigstar$) is statistically distinct from the invariable plane and the Neptune plane at greater than 99.7\% confidence.
According to \textsc{sct25} statistics (third block, second row), the Plutino mean plane is distinct from the Neptune plane at greater than 95\% confidence but not from the invariable plane.
According to \textsc{vm17} statistics (third block, third row), the Plutino mean plane is distinct from the invariable plane at greater than 95\% confidence and from the Neptune plane at greater than 99.7\% confidence.
All three methods agree that the local Laplace plane for the nominal semimajor axis of the exact 3:2 resonance (magenta $+$) is distinct at very high confidence.
The vMF mean plane is distant from the \textsc{sct25} and \textsc{vm17} mean planes, such that there is virtually no overlap in their 95\% ellipses, but the \textsc{sct25} and \textsc{vm17} methods produce similar mean plane estimates with similarly-sized 95\% ellipses with considerable overlap.
That the vMF method yields results largely deviant from the other two methods is unsurprising, as it is known that the observed sample of TNOs suffers from significant observational selection effects due to selection of observational survey fields \citep[e.g.][]{Gladman:2021}; the vMF method does not account for these biases, whereas the other two methods do.

Based on the presently known Plutino sample, the \textsc{sct25} method yields a precision of about 2 degrees while the \textsc{vm17} method yields a slightly lower precision, 2.59 degrees, for the location of the unbiased mean plane.
For the current observational sample sizes, the \textsc{sct25} method is more computationally practical than the \textsc{vm17} method and is more accurate than the vMF method, but as the observed Plutino population grows with future surveys it may become too large for the optimization step to handle, rendering the \textsc{vm17} method the better choice.
Below we report more detailed results for the Plutinos and their dynamical subgroups using the \textsc{sct25} method.

\section{The cataloged Plutinos, their $g$-librating subsets, and their \textsc{sct25} mean planes}
\label{s:Plutinos}
We take the list of observed Plutinos from \citet{volk2024dynamical}.
They apply a classification pipeline based on \citet{gmv08} to a catalog of 3357 TNOs and Centaurs from the Minor Planet Center and report 453 objects in Neptune's outer 3:2 MMR.
(Note that the paper specifies 452 Plutinos, but the accompanying data file lists 453, including Pluto.)
In addition, following \cite{malhotra2025doubly}, we identify the doubly resonant Plutinos, that is, those exhibiting libration of their argument of perihelion, $g$. (The $g$ librations are associated with the von Zeipel-Lidov-Kozai phenomenon, also known as the Kozai resonance or the Lidov-Kozai resonance, in the secular three-body problem \citep{Ito:2019,Tremaine:2023}.) 
\citet{malhotra2025doubly} examined the 453 Plutinos classified by \citet{volk2024dynamical} and identified 441 of them as stable residents of the 3:2 MMR over 100 Myr, with 69 of them as persistent $g$ librators over 100 Myr. 
Because the TNO classification pipeline of \citet{volk2024dynamical} uses a 10 Myr integration, we adopted the $g$-libration classification pipeline from \citet{malhotra2025doubly}, but with a 10 Myr integration; thus the sample sizes of the Plutinos and of the doubly resonant Plutinos are larger than those defined for stability over 100 Myr in \citet{malhotra2025doubly}.
This resulted in 446 stable Plutinos, 71 of which librate around $g=+90^\circ$ and 66 of which librate around $g=-90^\circ$.
We call these the $g+$ and $g-$ subgroups of Plutinos.
We note that, notionally, we can expect the unbiased populations of these subgroups to satisfy the condition of azimuthal symmetry about a mean plane because the vZLK dynamics does not restrict the orbit pole azimuth; however, because of the dynamical exchange between eccentricity and inclination in the vZLK libration, the \textsc{sct25} method's assumption that the inclination distribution is independent of the total angular momentum may not, \textit{a priori}, be expected to be satisfied by the vZLK librating source populations.
Therefore, the \textsc{sct25} estimates of the unbiased $g+$ and $g-$ mean planes must be interpreted with caution.

\begin{figure*}
\includegraphics[width=3in]
{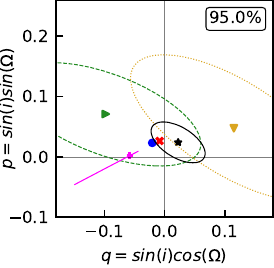}
\caption{
Mean plane estimates and 95\% credible regions (using \textsc{sct25}) for the doubly resonant subsets of the observed Plutinos:
the $g-$ group (yellow $\blacktriangledown$), 
the $g+$ group (green {\tiny $\blacktriangleright$}),
and mean plane for all Plutinos (black $\bigstar$).
We also indicate the following planes for comparisons:
the invariable plane (red $\times$),
Neptune's plane (blue $\bullet$),
the Laplace plane (magenta $+$) and its range over the Plutinos' range of semimajor axis (magenta line segment).
The reference frame is the J2000 ecliptic-equinox.
}
\label{fig:gplusminusnone_observed}
\end{figure*}

\begin{centering}
\begin{figure*}
\begin{subfigure}{.33\textwidth}
  \centering
  \includegraphics[width=.8\linewidth]{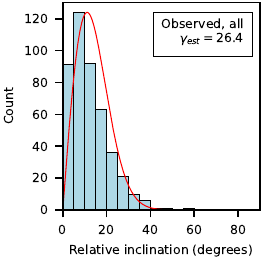}
  \label{fig:inclination_histograms_observed_gplusminusnone}
\end{subfigure}%
\begin{subfigure}{.33\textwidth}
  \centering
  \includegraphics[width=.8\linewidth]{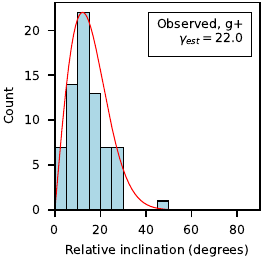}
  \label{fig:inclination_histograms_observed_gplus}
\end{subfigure}%
\begin{subfigure}{.33\textwidth}
  \centering
  \includegraphics[width=.8\linewidth]{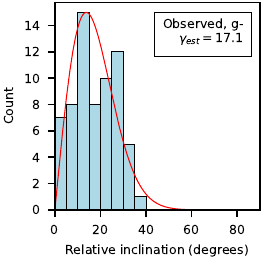}
  \label{fig:inclination_histograms_observed_gminus}
\end{subfigure}%
\caption{Inclination histograms of the observed Plutinos and their $g+$ and $g-$ subgroups;
these are measured relative to their respective \textsc{sct25} mean planes.
The red curve is the
vMF relative inclination distribution function that uses the parameter $\gamma_{\rm{est}}$ accompanying the \textsc{sct25} mean plane, as indicated in the legend.
}
\label{fig:inclination_histograms_observed}
\end{figure*}
\end{centering}

\begin{table}
    \begin{tabular}{lrrrrrlll}
    \hline\hline\addlinespace[3pt]
    \noalign{\hglue-4.5truein{Mean plane locations}} \addlinespace[1pt]
    & $n$ & $q$ & $p$ & $i$ & $\Omega$ & $\theta_{95\%}$ & $\gamma_{\rm{est}}$ & $\sigma_{\rm{est}}$ \\ \addlinespace[1pt]
    All & 446 & 0.022  & 0.024 & $1.88^\circ$ & $46.83^\circ$  & $2.04^\circ$ & 26.4 & $11.2^\circ$ \\
    $g+$ & 71 & -0.098 & 0.071 & $6.93^\circ$ & $144.06^\circ$ & $6.04^\circ$ & 22.0 & $12.2^\circ$ \\
    $g-$ & 66 & 0.115  & 0.047 & $7.13^\circ$ & $22.39^\circ$  & $7.01^\circ$ & 17.1 & $13.8^\circ$ \\ \addlinespace[3pt]
    \hline \addlinespace[3pt]
    \noalign{\hglue-3.7truein{Mean plane separations (in degrees)}} \addlinespace[1pt]
          & $n$ & All  & $g+$ & $g-$ & Invariable & Neptune & Laplace \\ \addlinespace[1pt]
        All  & 446 & 0    & 7.41  &  5.47 & 1.77 & 2.46 & 4.76 \\
        $g+$ & 71  & 7.41 & 0     & 12.27 & 5.74 & 5.22 & 4.58 \\
        $g-$ & 66  & 5.47 & 12.27 & 0     & 7.17 & 7.89 & 10.24 \\ \addlinespace[3pt]
        \hline \addlinespace[3pt]
         \noalign{\hglue-5.1truein{$P$-values}} \addlinespace[1pt]
          & $n$ &  &  &  & Invariable & Neptune & Laplace \\ \addlinespace[1pt]
        All  & 446 &    &  &  & 0.17  & 0.018  & 0 \\
        $g+$ & 71  &  &    &  & 0.32  & 0.35   & 0.12 \\
        $g-$ & 66  &  &  &    & 0.020 & 0.0084 & 0.00012 \\ \addlinespace[3pt]
        \hline\hline
    \end{tabular}
    \caption{
Data accompanying Figure \ref{fig:gplusminusnone_observed}.
The first block reports the locations of the mean planes of all the observed Plutinos, and for their $g+$ and $g-$ librating subsets;
the first column tabulates the sample size and the last columns tabulate the concentration parameter of the best-fit vMF function for the inclinations relative to the mean plane and the corresponding width of the Rayleigh relative inclination distribution.
The second block reports the angular separations amongst the three mean planes, as well as their separations from three physically relevant planes (the invariable plane, Neptune’s plane and the Laplace plane).
The third block reports the $P$-values of the angular separations of the estimated mean planes from the three reference planes.
See main text for details.
}
\label{t:figure2_meanplanes}
\end{table}

Figure \ref{fig:gplusminusnone_observed} shows the debiased mean planes of all observed Plutinos and of their $g+$ and $g-$ librating subsets, according to \textsc{sct25} statistics.
The accompanying mean plane locations, mean plane separations (from each other and from three reference planes), and the associated statistical $P$-values are listed in Table \ref{t:figure2_meanplanes}.

First, we comment on the locations of the mean planes relative to the three physically relevant planes, i.e., Neptune's plane, the Laplace plane and the invariable.
The mean plane of all Plutinos is statistically distinct from  Neptune's plane at the 95\% level and from the Laplace plane at the 99.7\% level, but not from the invariable plane.
The mean plane of the $g-$ subset is statistically distinct from Neptune's plane at close to the 99.7\% level, from the Laplace plane at better than the 99.7\% level, and from the invariable plane at the 95\% level.
The mean plane of the $g+$ subset is not statistically distinct from Neptune's plane, the Laplace plane or the invariable plane.

Next, we consider the relative locations of the measured mean planes.
The angular distances amongst the mean planes of the three groups are strikingly large: $7.41^\circ$ between the $g+$ mean plane and the mean plane of all Plutinos, $5.47^\circ$ between the $g-$ mean plane and the mean plane of all Plutinos, and $12.27^\circ$ between the $g+$ and $g-$ mean planes, with ecliptic mean inclinations of $5.74^\circ$ for the $g+$ subset and $7.17^\circ$ for the $g-$ subset.
However, the mean planes of the $g+$ and $g-$ groups have large credibility regions owing to their smaller sample sizes and their larger overall inclinations.
Consequently, the mean plane of all Plutinos (black $\bigstar$) lies within the 95\% probability contours for both the $g+$ subset (green {\tiny $\blacktriangleright$}) and the $g-$ subset ($\blacktriangledown$).

Despite their relatively large credible regions, the mean planes of the two $g$-librating subsets each lie outside the other's 99.7\% credibility contour.
(It should be understood that this visual observation is not a statistical comparison of means, so we do not associate it with a $P$-value.)
We note that the $g+$ and $g-$ mean planes appear distinct not only in their inclination, but also in their longitude of the ascending node.
This is unexpected, because the vZLK resonance acts on eccentricity, inclination, and argument of pericenter but no associated asymmetries in the longitude of ascending node have been noted in the literature.
On the $(q,p)$ plane, we observe that the $g+$ and $g-$ mean planes are nearly collinear with the mean plane of the Plutinos, with the latter located in-between that of the $g+$ and $g-$ subgroups.
The angle between the $g+$ and $g-$ mean plane locations on the $(q,p)$ plane, measured with the mean plane of all Plutinos as the vertex, is $144^\circ$. 
That is, the $g+$ and $g-$ mean planes are separated by $\Omega_{\rm{rel}}=144^\circ$ of their longitudes of ascending node measured on the mean plane of all Plutinos.
This value is intriguingly close to $180^\circ$, the angular difference in $g$ libration centers of the two groups.
Due to the large uncertainty ellipses of the $g+$ and $g-$ mean planes, we consider this a tentative result.
We leave to future analysis of larger samples to ascertain and explain this result.

Figure \ref{fig:inclination_histograms_observed} plots histograms of the inclination of the observed Plutinos and their $g+$ and $g-$ subsets, measured relative to their \textsc{sct25} mean planes.
Also superimposed is the best-fit vMF function \citep{mmk23}, as determined by the \textsc{sct25} method.
In each case the debiased vMF function indicates an intrinsic distribution of inclinations that is somewhat wider than their apparent inclination distribution.
Notably, the debiased vMF concentration parameter for all Plutinos from the \textsc{sct25} method is $\gamma_{\rm{est}}=26.4$, compared to $\gamma_{\rm{est}}=31.5$ for the biased (apparent) vMF estimate in Table \ref{t:figure1_meanplanes}.
In other words, the \textsc{sct25} method finds that the Plutinos' intrinsic inclination distribution is somewhat wider than their apparent distribution; this is consistent with expectations, as the observational surveys are biased towards low-ecliptic latitudes, hence lower inclinations overall.
Likewise, the inclinations of the $g+$ and $g-$ subgroups are also more widely dispersed than their apparent distributions, and even somewhat wider than those of the Plutinos as a whole.
Approximated as Rayleigh distributions, the inclination dispersions of all Plutinos and the $g+$ and $g-$ subgroups have widths of 11.2, 12.2 and 13.8 degrees, respectively.

It is also widely separated from the Laplace plane (i.e., the linear Laplace-Lagrange secular theory estimate of the forced plane about which orbits precess). 
We speculate that this result is due to the 3:2 MMR's proximity to the $\nu_{18}$ secular resonance. 
This resonance warps the Laplace plane at $a\approx40.3$ au, just outside the semimajor axis range of the stable Plutinos. 
With the Laplace plane itself singular at that location, the next convenient hypothesis is that over long timescales the average orbital angular momentum of the Plutinos will point in the same direction as that of the Solar System at large, or the invariable plane.

As a point of comparison to the observed Plutino population, we examined the mean plane of the large synthetic Plutino sample of \cite{lawler2025exploring}.
In that work, the sample was generated to fill uniformly the orbital parameter space of Neptune's 3:2 mean motion, with semimajor axes in the range 39.0--39.8 au, eccentricities up to 0.5 and ecliptic inclinations up to 90$^\circ$. 
The authors reported that after a 4 gyr integration, 69,626 particles remained librating stably in the resonance; this is the synthetic sample that we analyze (details described in Appendix \ref{sec:appendix_lawler}).  
For this synthetic sample, we find that the mean plane of the complete set has a 95\% probability mass contour of angular size $\theta_{95\%}=0.37^\circ$ and is distinct from the Laplace plane and from Neptune's plane at greater than 99.7\% confidence but consistent with the invariable plane at 95\% confidence.
The mean planes of the $g+$ and $g-$ subsets are located outside the 95\% probability contour of the mean plane of the complete set, although their angular distances to the mean plane of the complete set are 0.63 and 0.65 degrees, respectively, much smaller that the corresponding distances for the observed doubly resonant Plutino samples.
We notice a qualitatively similar pattern as for the observed Plutinos that we discussed above, namely, 
on the $(q,p)$ plane, the $g+$ and $g-$ mean planes are nearly collinear with the mean plane of the complete set, with the latter located in-between that of the $g+$ and $g-$ subgroups.
The angle between the $g+$ and $g-$ mean plane locations on the $(q,p)$ plane, measured 
with the mean plane of the complete set as the vertex, is $142^\circ$, similar to 144$^\circ$ for the observed Plutinos.
Note, however, that for the observed and simulated Plutinos these angles are measured in opposite senses: the observed $g+$ mean plane is west of the observed $g-$ mean plane, and the simulated $g+$ mean plane is east of the simulated $g-$ mean plane.
In contrast with the observed Plutinos, the distribution of inclinations relative to the mean planes are much more widely dispersed and not well described with a vMF function; this suggests that the initialization of the sample is not congruent with the origins of the real Plutino population (as noted in \cite{lawler2025exploring}).

\section{The mean planes of other Kuiper belt MMRs}
\label{sec:appendix_mmrs} 

\begin{centering}
\begin{figure*}
\begin{subfigure}{.25\textwidth}
  \centering
  \includegraphics[width=\linewidth]{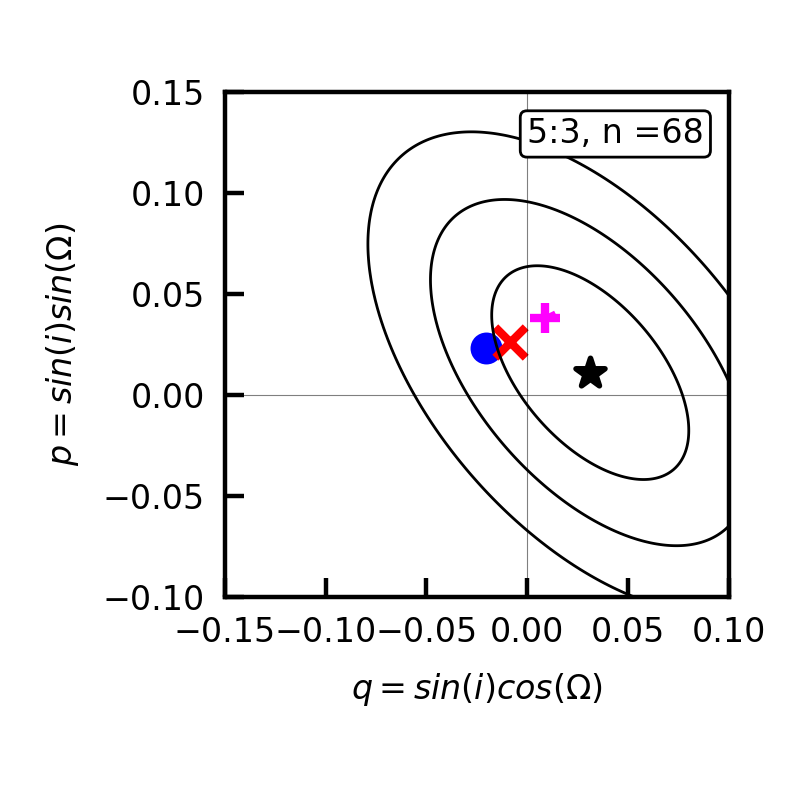}
  \label{fig:plot_contours_mmrs_53}
\end{subfigure}%
\begin{subfigure}{.25\textwidth}
  \centering
  \includegraphics[width=\linewidth]{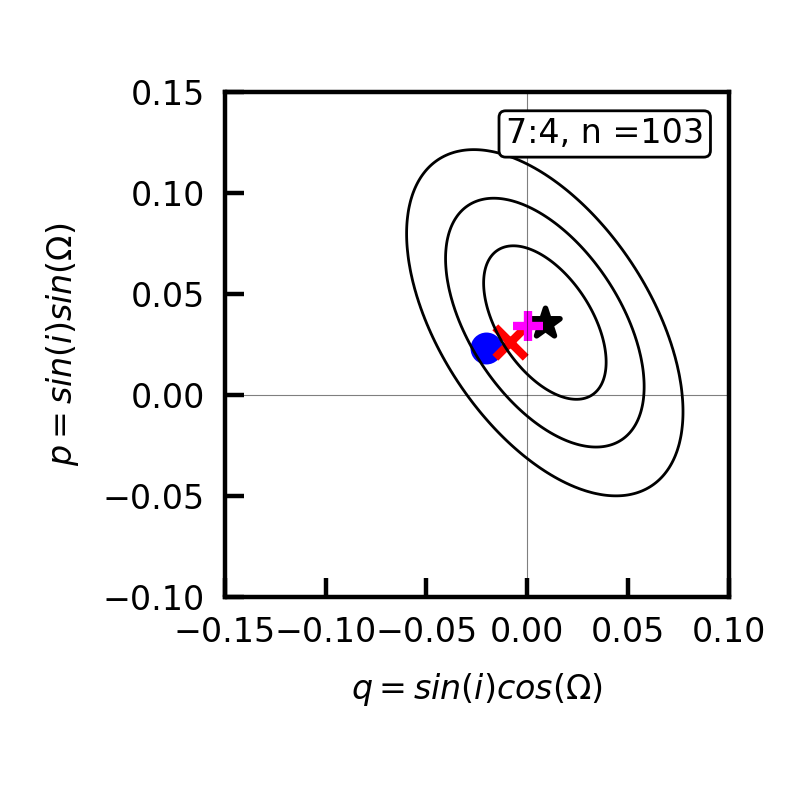}
  \label{fig:plot_contours_mmrs_74}
\end{subfigure}%
\begin{subfigure}{.25\textwidth}
  \centering
  \includegraphics[width=\linewidth]{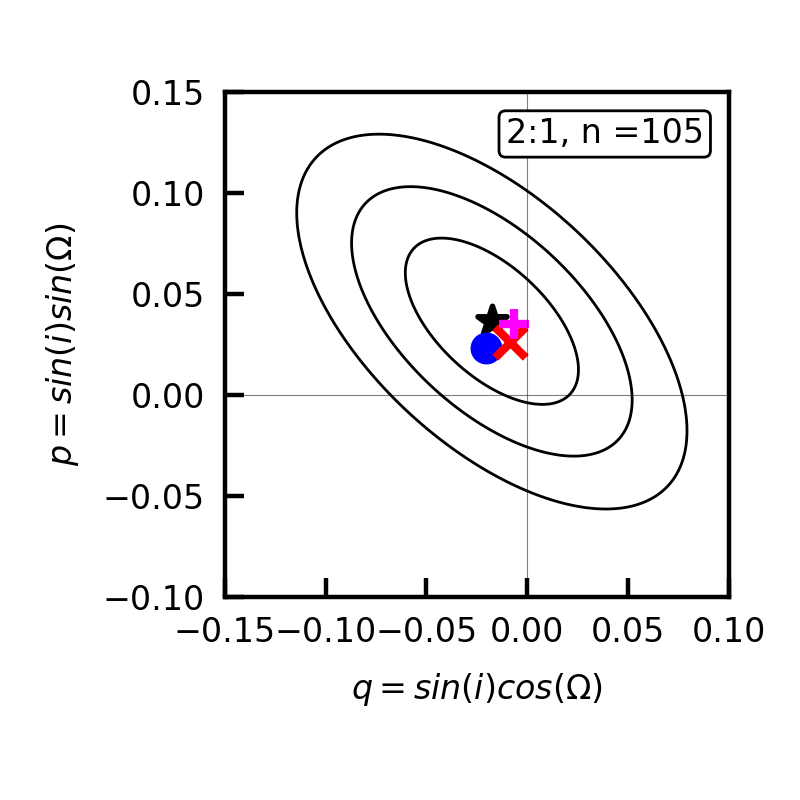}
  \label{fig:plot_contours_mmrs_21}
\end{subfigure}%
\begin{subfigure}{.25\textwidth}
  \centering
  \includegraphics[width=\linewidth]{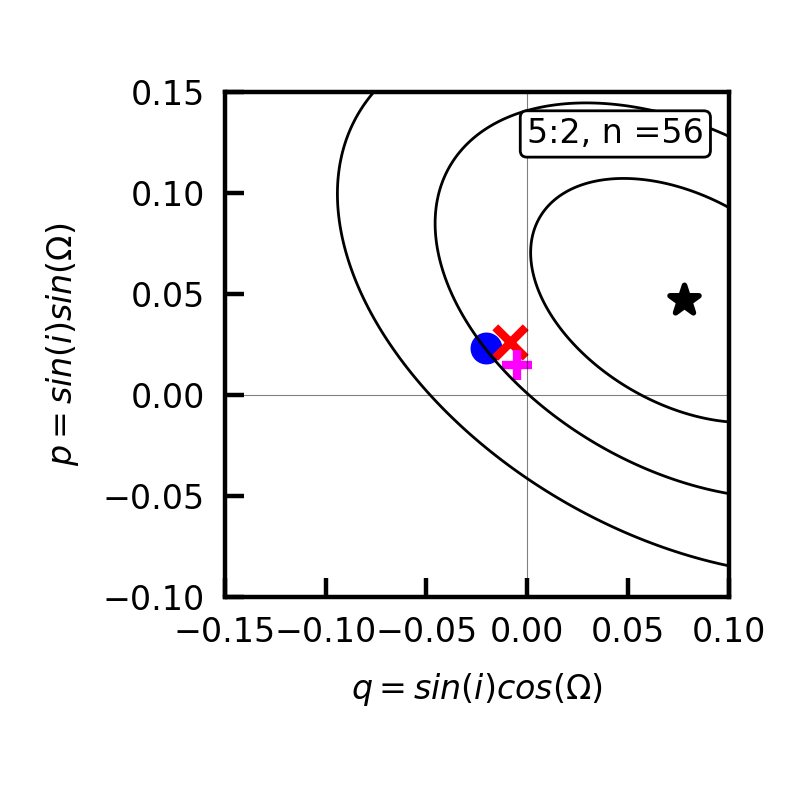}
  \label{fig:plot_contours_mmrs_52}
\end{subfigure}%
\caption{\textsc{sct25} mean planes (black $\bigstar$) and 68\%, 95\%, and 99.7\% credible ellipses for the objects in Neptune's four most populous resonances besides the Plutinos. 
Reference planes include the invariable plane (red $\times$), Neptune plane (blue $\bullet$), and local Laplace plane (magenta $+$).
The reference frame is the J2000 ecliptic-equinox.}
\label{fig:plot_contours_mmrs}
\end{figure*}
\end{centering}

We used the \textsc{sct25} method to calculate the mean plane and mean plane credible regions for the populations in Neptune's outer 5:3, 7:4, 2:1, and 5:2 MMRs, with respective population sizes of 68, 103, 105, and 56 objects drawn from the list of TNOs classified by \citet{volk2024dynamical}.
These are plotted in Figure \ref{fig:plot_contours_mmrs}.
As for the Plutinos, we examined the $g$ libration status of these objects over a 10 Myr integration.
We found 2, 9, 3, and 8 $g+$ librators in each respective MMR and 4, 8, 1, and 1 $g-$ librator, respectively.
We judged these $g$-librating populations too small to compute their mean planes and mean plane credible regions separately.
Table \ref{t:figure6_meanplanes} contains the mean plane locations of the MMRs, the average angular widths of their 95\% credible ellipses; their angular separations from the invariable plane, Neptune orbital plane, and Laplace plane; and the $P$-values for those separations.

At $1.90^\circ$, $2.12^\circ$, and $2.32^\circ$, the respective ecliptic inclinations of mean planes for the 5:3, 7:4, and 2:1 MMRs are slightly higher than the $1.88^\circ$ ecliptic  inclination of the Plutinos' mean plane, and the $5.20^\circ$ inclination of the 5:2 MMR mean plane is higher still.
Although these populations have (mostly) narrower inclination widths of $9.5^\circ$, $7.8^\circ$, $9.4^\circ$, and $12.4^\circ$ compared to the $11.2^\circ$ inclination width of the Plutinos, their smaller population sizes produce larger uncertainty intervals in their mean plane locations than for the Plutinos.
The mean plane of the Plutinos can be located with a precision of $2.04^\circ$, but the other MMR mean planes can only be located with precisions ranging from $2.92^\circ$ to $6.03^\circ$.
Hence, the other mean planes do not appear statistically distinct from each other at the 95\% level, and they are all consistent with Neptune's plane, the local Laplace plane, and the invariable plane at or about the 95\% level.

\begin{centering}
\begin{figure*}
\begin{subfigure}{.25\textwidth}
  \centering
  \includegraphics[width=\linewidth]{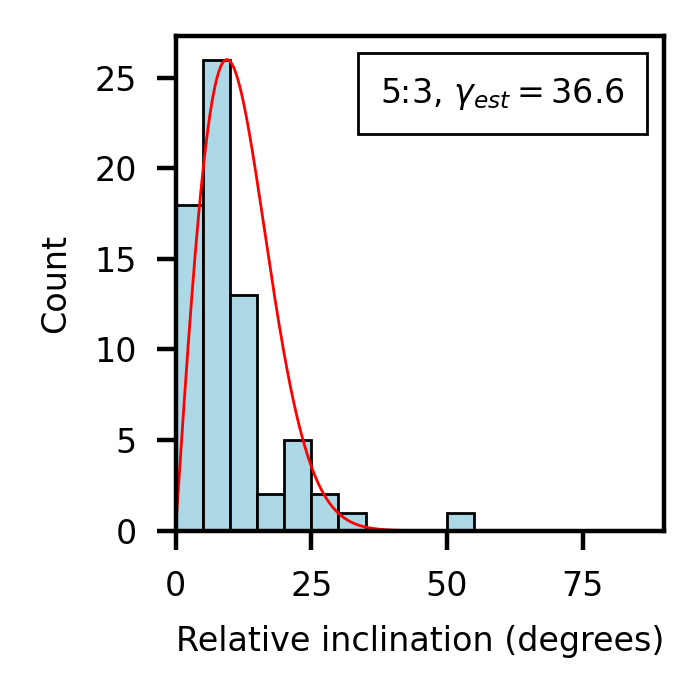}
  \label{fig:inclination_histograms_mmrs_53}
\end{subfigure}%
\begin{subfigure}{.25\textwidth}
  \centering
  \includegraphics[width=\linewidth]{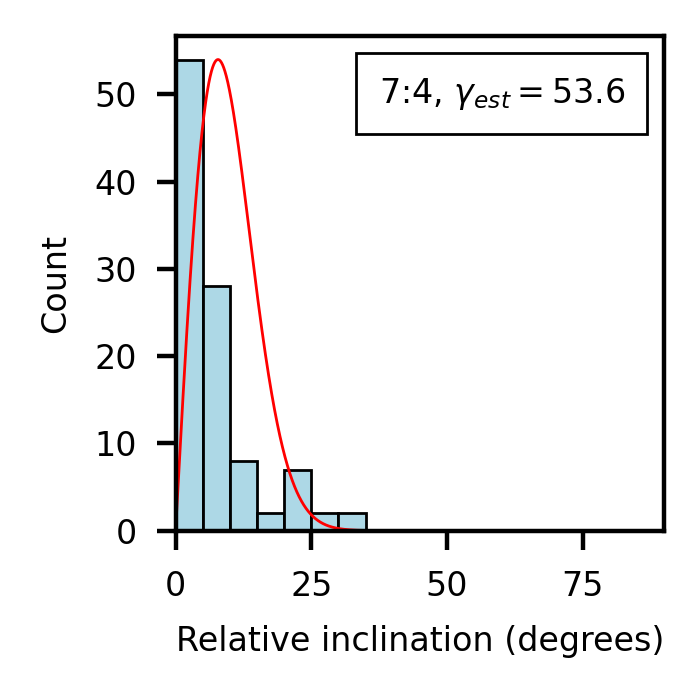}
  \label{fig:inclination_histograms_mmrs_74}
\end{subfigure}%
\begin{subfigure}{.25\textwidth}
  \centering
  \includegraphics[width=\linewidth]{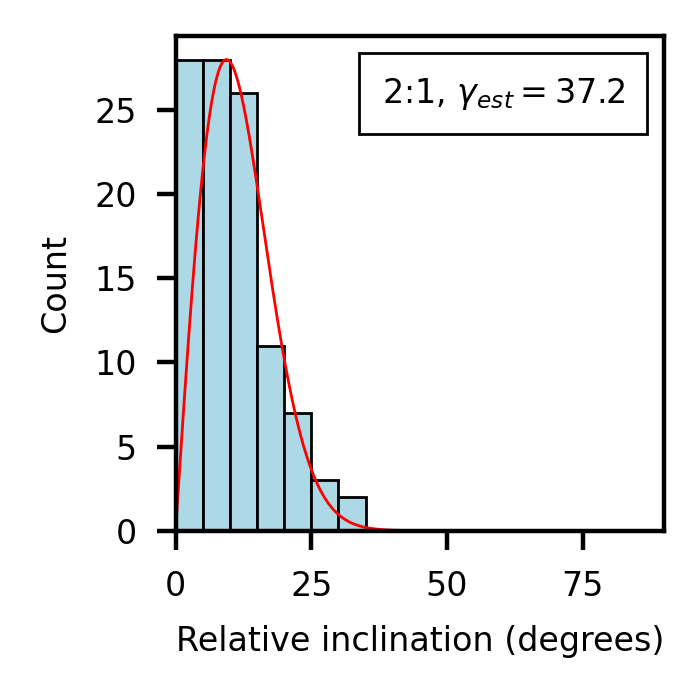}
  \label{fig:inclination_histograms_mmrs_21}
\end{subfigure}%
\begin{subfigure}{.25\textwidth}
  \centering
  \includegraphics[width=\linewidth]{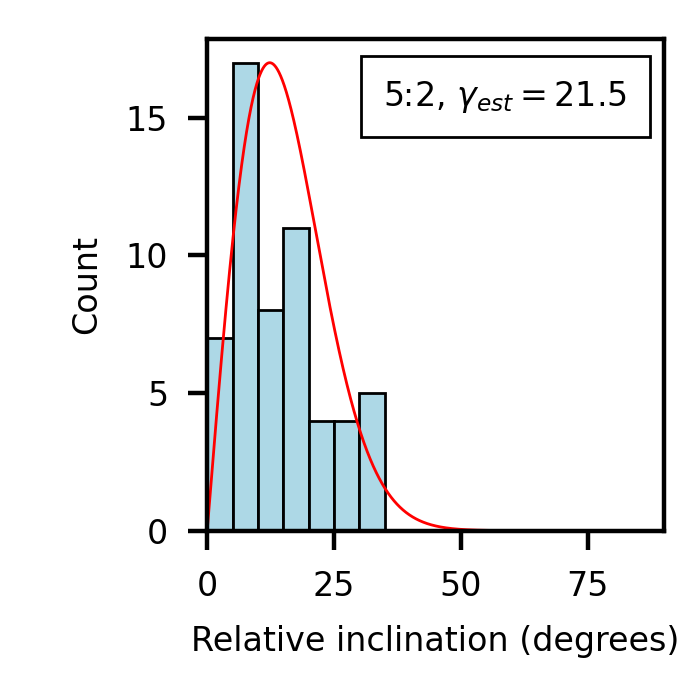}
  \label{fig:inclination_histograms_mmrs_52}
\end{subfigure}%
\caption{
Inclination histograms of the observed objects in Neptune's four most populous resonances besides the Plutinos;
these are measured relative to their respective \textsc{sct25} mean planes.
The red curve is the
vMF relative inclination distribution function with parameter $\gamma_{\rm{est}}$, as found by the \textsc{sct25} maximum-likelihood mean plane computation.  
Note that each panel has its own scale on the ordinate.}
\label{fig:inclination_histograms_mmrs}
\end{figure*}
\end{centering}

\begin{table}
    \begin{tabular}{lrrrrrlll}
    \hline\hline\addlinespace[2pt]
    \noalign{\hglue-4.5truein{Mean plane locations}} \addlinespace[1pt]
    & $n$ & $q$ & $p$ & $i$ & $\Omega$ & $\theta_{95\%}$ & $\gamma_{\rm{est}}$ & $\sigma_{\rm{est}}$ \\ \addlinespace[1pt]
    5:3 &  68 &  0.031  & 0.011 & $1.90^\circ$ &  $19.41^\circ$ & $4.34^\circ$ & 36.6 & $9.5^\circ$ \\
    7:4 & 103 &  0.0087 & 0.036 & $2.12^\circ$ &  $76.30^\circ$ & $2.92^\circ$ & 53.6 & $7.8^\circ$ \\
    2:1 & 105 & -0.018  & 0.037 & $2.32^\circ$ & $115.64^\circ$ & $3.52^\circ$ & 37.2 & $9.4^\circ$ \\
    5:2 &  56 &  0.078  & 0.047 & $5.20^\circ$ &  $31.07^\circ$ & $6.03^\circ$ & 21.5 & $12.4^\circ$ \\ \addlinespace[3pt]
    \hline \addlinespace[3pt]
    \noalign{\hglue-3.7truein{Mean plane separations (in degrees)}} \addlinespace[1pt]
          & $n$ & Invariable & Neptune & Laplace \\ \addlinespace[1pt]
        5:3  &  68 & 2.43 & 3.05 & 2.01  \\
        7:4  & 103 & 1.12 & 1.83 & 0.49  \\
        2:1  & 105 & 0.79 & 0.79 & 0.62  \\
        5:2  &  56 & 5.07 & 5.80 & 5.09  \\ \addlinespace[3pt]
        \hline \addlinespace[3pt]
         \noalign{\hglue-5.1truein{$P$-values}} \addlinespace[1pt]
          & $n$ & Invariable & Neptune & Laplace \\ \addlinespace[1pt]
        5:3  &  68 & 0.46 & 0.23 & 0.71  \\
        7:4  & 103 & 0.44 & 0.12 & 0.87  \\
        2:1  & 105 & 0.92 & 0.79 & 0.91  \\
        5:2  &  56 & 0.10 & 0.05 & 0.08  \\ \addlinespace[3pt]
        \hline\hline
    \end{tabular}
    \caption{
Data accompanying Figure \ref{fig:plot_contours_mmrs}.
The first block reports the locations of the mean planes of Neptune's four most populous MMRs besides the Plutinos;
the first column tabulates the sample size and the last columns tabulate the concentration parameter of the best-fit vMF function for the inclinations relative to the mean plane and the corresponding width of the Rayleigh relative inclination distribution.
The second block reports the angular separations amongst the three mean planes, as well as their separations from three physically relevant planes (the invariable plane, Neptune’s plane and the Laplace plane).
The third block reports the $P$-values of these angular separations.
See main text for details.
}
\label{t:figure6_meanplanes}
\end{table}

\begin{figure}
\includegraphics[width=0.8\textwidth]
{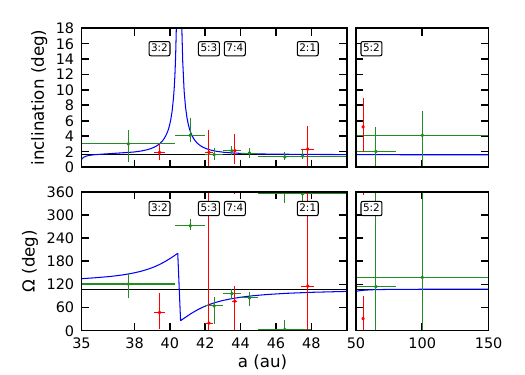}
\caption{
\textsc{sct25} mean planes and 68\% confidence intervals for the 3:2, 5:3, 7:4, 2:1, and 5:2 MMRs are plotted in red (this work), the mean planes and 68\% confidence intervals for various non-resonant Kuiper belt semimajor axis bins are plotted in green (from \citet{mm23}); the Laplace plane is indicated by the blue curves.
The reference frame is the J2000 ecliptic-equinox.
}
\label{fig:iw_vm17plot}
\end{figure}

\section{Summary and Discussion}
\label{s:summary}
We computed the unbiased mean plane of the Plutinos as well as those of their $g$-librating subsets whose arguments of pericenter librate around either $g=+90^\circ$ or $g=-90^\circ$ (also known as the doubly resonant Plutinos or the vZLK-librators).
We also estimated the inclination distribution about these mean planes.
In addition to the Plutinos, we computed the mean planes and inclination distributions of four other resonant Kuiper belt groups, those in the 5:3, 7:4, 2:1 and 5:2 MMRs.
These analyses used the unbiased methods described in \citet{siraj2025measuring} and in \citet{vm17} and are based on the current observational samples; these two methods yield substantially similar results.
Below, we summarize the results obtained with the method of \citet{siraj2025measuring}.

\begin{enumerate}
\item
The mean plane of the Plutinos is located at $i=1.88$ degrees and $\Omega=46.83$ degrees in the J2000 reference frame, with a 95\% credible error of up to $2.04$ degrees from its calculated location. 
It is statistically indistinguishable from the invariable plane of the Solar System, but quite distant from the local Laplace plane.
\item 
The $g+$ and $g-$ librators amongst the Plutinos have mean plane locations that are uncertain by about $6^\circ$ and $7^\circ$, respectively, at 95\% credibility.
Despite these significant uncertainties, these two dynamical subgroups' unbiased mean planes appear different from each other at the 99.7\% level, with mutual inclination of $12.27^\circ$ and separated by $144^\circ$ in the longitude of ascending node when $\Omega$ is reckoned relative to the mean plane of the entire population. 
We consider this a tentative result owing to an important caveat, namely that the source populations of these subgroups may not satisfy the assumption of the \textsc{sct25} method that their inclination distribution is separable from their orbital energy and total angular momentum distribution.
\item The 5:3, 7:4, 2:1, and 5:2 MMRs each have mean plane locations that are statistically indistinguishable at or about the 95\% level from the invariable plane, the local Laplace plane, and the orbital plane of Neptune. 
However, their mean plane locations are presently measurable with a precision of only about $3^\circ$--$6^\circ$, at 95\% credibility.
\item
For all resonant populations, the inclinations relative to their measured mean planes are fairly well represented with the von Mises-Fisher function. The inclination dispersions of all Plutinos and their $g+$ and $g-$ subgroups have widths of 11.2, 12.2 and 13.8 degrees, respectively. 
The inclination dispersions in Neptune's 5:3, 7:4, 2:1, and 5:2 MMRs have widths of 9.5, 7.8, 9.4 and 12.4 degrees, respectively.
\end{enumerate}

As a summary, in Figure \ref{fig:iw_vm17plot} we plot the ecliptic inclination and longitude of ascending node of all the mean planes of resonant Kuiper belt groups computed in this work, together with the mean planes of non-resonant Kuiper belt objects computed in previous work \citep{mm23}. 
We observe that, overall, the resonant mean planes (red) have inclination uncertainties similar to those of the non-resonant mean planes (green), although the uncertainties in the longitude of ascending node are somewhat larger.
Except for the Plutinos, the mean planes across all measured semimajor axes ranges are consistent with the local Laplace plane (the linear Laplace-Lagrange secular theory estimate of the local forced plane), and those of the other resonant groups are nearly the same compared to the mean planes of nearby non-resonant populations.

Our result for the mean plane of the Plutinos is in some contrast with what we found in our recent study of the Hilda asteroids, the group of small bodies in Jupiter's 3:2 interior MMR \citep{mm26hildas}.
The mean plane of the Hildas is distinct from the invariable plane but is statistically consistent with the local Laplace plane and with Jupiter's plane at 95\% confidence, whereas we find that the mean plane of the Plutinos is quite distinct from the local Laplace plane but statistically indistinguishable from the invariable plane.
We note that the precision of the Hildas' mean plane is nearly an order-of-magnitude better than that of the Plutinos' mean plane, therefore it is not possible to make strong conclusions based on these contrasting results.
It is also worth noting that there are two significant differences in the environments of these two resonant populations:
the perturbing planet's mass is much larger for the Hildas,
and the Plutinos are proximate to the $\nu_{18}$ secular resonance which strongly warps the local Laplace plane near $\sim40$~au, just outside the semimajor axis range of the Plutinos (see Fig.~\ref{fig:iw_vm17plot}).
Moreover, the Plutinos are affected by the vZLK oscillation, whereas we find no vZLK librators amongst the Hildas.
These results underscore the complexity of the spatial dynamics of mean motion resonances.
More analysis is needed to understand the effects of resonant perturbations on the forced planes of small body populations, and to understand how mean motion resonances interact with secular dynamics and with the vZLK oscillation to produce distinct forced planes for different $g$ libration centers.

Semi-analytical theory with orbit-averaging techniques has been used in \citet[][and references therein]{malhotra2025doubly} to clarify some aspects of the vZLK dynamics in a mean motion resonance.
These studies have modeled multiple planetary perturbers in circular, co-planar orbits.
Development of the theory along these lines, to account for non-coplanar planetary orbits, could potentially help to calculate the forced plane of a small body in a mean motion resonance, and provide predictions for testing with observations.
Future work could also consider validation and/or extension of the \textsc{sct25} method for dynamical populations in which the inclination distribution is not separable from the distribution of the orbital energy and total angular momentum, such as the vZLK Plutinos or even some non-resonant vZLK populations.

The upcoming LSST survey by the Rubin Observatory is expected to make well-characterized detections of up to 5220 Plutinos, or 11.7 times as many as in today's observed population \citep{kurlander2025predictions}.
With such a large sample size, we can estimate that the statistical credible regions of the mean plane locations would decrease by a factor of about $\sqrt{11.7}\approx3.4$.
This increase in measurement precision offers the prospect of better distinguishing the observed mean planes of the Plutinos and of their vZLK subsets from the invariable plane and other reference planes, to potentially identify unmodeled effects of undiscovered distant perturbers, or to reconcile the mean planes with one of the reference planes.
Future work could improve our understanding of LSST's potential to resolve the Plutinos' mean plane by examining the mean plane distribution of an ensemble of simulated LSST Plutino detections.

\section{Data Availability}
\label{s:data_availability}
The initial orbital elements for the planets and KBOs, as well as our Python code, are available for download from \url{https://github.com/iwygh/mm26_Plutinos}.

\section{Acknowledgments}
We thank 
Amir Siraj for correspondence about how to implement the \textsc{sct25} mean plane calculation, 
Samantha Lawler for correspondence regarding the simulations in \citet{lawler2025exploring}, 
and the anonymous reviewers for detailed comments that helped to improve this paper.
Funding for IM was provided by NASA FINESST grant 80NSSC23K1362.
This material is based upon High Performance Computing (HPC) resources supported by the University of Arizona TRIF, UITS, and Research, Innovation, and Impact (RII) and maintained by the UArizona Research Technologies department.
This research has made use of data and/or services provided by the International Astronomical Union's Minor Planet Center, and of the Horizons service provided by JPL's Solar System Dynamics Group.
We have made extensive use of the Python packages \texttt{Astroquery} \citep{2019AJ....157...98G} and \texttt{Rebound} \citep{rebound}.

\printcredits

\bibliographystyle{cas-model2-names}

\bibliography{refs_Plutinos}


\appendix
\section{The mean plane of a synthetic Plutino population}\label{sec:appendix_lawler} 
We wish to compare the calculated mean planes of the observed Plutinos and their $g$-librating subsets, along with the statistical confidence in those calculations, to a theoretical prediction for the forced plane.
The question is how to define that prediction.
Linear secular theory offers an easily calculated forced plane location for the nominal central semimajor axis of the MMR, but its averaging over the mean motion of all bodies makes the prediction unsuitable within MMRs.
Higher-order predictions using the disturbing function with mean motion terms retained are analytically intractable.
Numerical simulations can reveal the mean plane that emerges over time from a model population, but that mean plane may be model-dependent and time-dependent,
and its relationship to dynamical forcing may not be immediately apparent.

As an expedient alternative to conducting a battery of simulations ourselves, we turn to the simulated Plutino sample from \citet{lawler2025exploring}.
In that study, one million test particles were seeded uniformly across $39.0<a<39.8$ au, $0<e<0.5$, and $0<i<90^\circ$, with $\Omega$, $\omega$, and $M$ seeded uniformly on the circle
and $i$, $\Omega$, and $\omega$ defined relative to the J2000 ecliptic-equinox reference frame.
The test particles were integrated with the four giant planets for 10 Myr and then for longer durations, dropping the objects that did not remain stable in the 3:2 MMR at each stage until, after 4 Gyr, only 62,097 Plutinos remained both stable and not librating in $g$.
The $g$-librating population comprised 3740 objects librating around $g=+90^\circ$ and 3789 around $g=-90^\circ$.
The total count of stable Plutinos outside the $g$ libration and Plutinos stably librating around either $g=+90^\circ$ or $g=-90^\circ$ was 69,626.
The proportion of $g$-librating Plutinos is only 10.8\% in the simulated sample, versus 30.7\% in the observed catalog.
In both samples, the $g+$ and $g-$ librators have populations of similar sizes: In the observed catalog there are 7.5\% more $g+$ librators than $g-$ librators, and in the simulated sample there are 1.3\% fewer $g+$ librators than $g-$ librators.

\citet{lawler2025exploring} consider the orbital element distribution of the 69,626 remaining objects to fill the parameter space of the 3:2 MMR.
That is, to a good approximation, every combination of orbital elements that can result in a stable Plutino is represented by one of the remaining objects.
It does not model the entire Plutino population as present today after several Gyr of evolution from the primordial planetary disk, but it does represent a maximally populated MMR with no empty regions.
Thus, the mean plane of this uniform simulated sample may be said to represent the mean plane 
(and perhaps the forced plane)
of Neptune's outer 3:2 MMR if it were only subject to gravitational perturbations from the outer planets as might be predicted by some analytical theory for the forced plane in a mean motion resonance.

\begin{figure*}
\includegraphics[width=3in]
{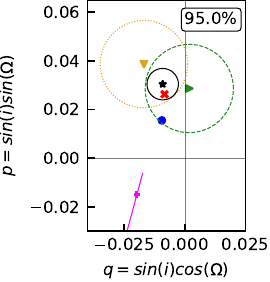}
\caption{
vMF mean planes and 95\% confidence circles for the synthetic Plutinos and their $g$-librating subsets, featuring the $g-$ mean plane (yellow $\blacktriangledown$), $g+$ mean plane (green {\tiny $\blacktriangleright$}), and mean plane for all stable Plutinos plus stable $g$ librators (black $\bigstar$).
Reference planes include the invariable plane (red $\times$), Neptune plane (blue $\bullet$), and Laplace plane (magenta $+$).
The magenta line segment is the range of the Laplace plane over the Plutinos' range of semimajor axis.
The reference frame is the J2000 ecliptic-equinox.
}
\label{fig:gplusminusnone_lawler}
\end{figure*}

\begin{table}
    \begin{tabular}{lrrrrrlll}
    \hline\hline\addlinespace[3pt]
    \noalign{\hglue-4.5truein{Mean plane locations}} \addlinespace[1pt]
    & $n$ & $q$ & $p$ & $i$ & $\Omega$ & $\theta_{95\%}$ & $\gamma_{\rm{est}}$ & $\sigma_{\rm{est}}$ \\ \addlinespace[1pt]
    All & 69,626 & -0.0090 & 0.030 & $1.82^\circ$ & $106.68^\circ$ & $0.37^\circ$ & 2.99 & $33.2^\circ$ \\
    $g+$ & 3,740 & 0.0020  & 0.029 & $1.64^\circ$ &  $86.34^\circ$ & $1.04^\circ$ & 6.14 & $23.1^\circ$ \\
    $g-$ & 3,789 & -0.017  & 0.039 & $2.41^\circ$ & $113.51^\circ$ & $1.03^\circ$ & 6.18 & $23.0^\circ$ \\ \addlinespace[3pt]
    \hline \addlinespace[3pt]
    \noalign{\hglue-3.7truein{Mean plane separations (in degrees)}} \addlinespace[1pt]
          & $n$ & All  & $g+$ & $g-$ & Invariable & Neptune & Laplace \\ \addlinespace[1pt]
        All  & 69,626 & 0    & 0.63 & 0.65 & 0.24 & 0.85 & 2.68 \\
        $g+$ & 3,740  & 0.63 & 0    & 1.21 & 0.60 & 0.99 & 2.79 \\
        $g-$ & 3,789  & 0.65 & 1.21 & 0    & 0.86 & 1.39 & 3.08 \\ \addlinespace[3pt]
        \hline \addlinespace[3pt]
         \noalign{\hglue-5.1truein{$P$-values}} \addlinespace[1pt]
          & $n$ &  &  &  & Invariable & Neptune & Laplace \\ \addlinespace[1pt]
        All  & 69,626 &  &  &  & 0.28 & $1.1\times10^{-7}$ & 0 \\
        $g+$ & 3,740  &  &      &   & 0.37 & 0.065 & 0 \\
        $g-$ & 3,789  & &   &      & 0.12 & 0.0041 & 0 \\ \addlinespace[3pt]
        \hline\hline
    \end{tabular}
    \caption{
Data accompanying Figure \ref{fig:gplusminusnone_lawler}.
The first block reports the locations of the mean planes of all the synthetic Plutinos, and of their $g+$ and $g-$ librating subsets;
the first column tabulates the sample size and the last columns tabulate the concentration parameter of the best-fit vMF function for the inclinations relative to the mean plane and the corresponding width of the Rayleigh relative inclination distribution.
The second block reports the angular separations amongst the three mean planes, as well as their separations from three physically relevant planes (the invariable plane, Neptune’s plane and the Laplace plane).
The third block reports the $P$-values of the angular separations of the estimated mean planes from the three reference planes.
See main text for details.
}
\label{t:figure4_meanplanes}
\end{table}

Because this synthetic Plutino population is not subject to observational bias, its mean plane and the mean planes of its $g$-librating subsets at the end of the 4 Gyr integration can be computed using simple vMF statistics.
These mean planes, and the reference Laplace plane, are computed in the same heliocentric J2000 ecliptic-equinox coordinates that the end states of the planets and synthetic Plutinos are given in.
Figure \ref{fig:gplusminusnone_lawler} shows those mean planes,
and Table \ref{t:figure4_meanplanes} displays the mean plane locations, their angular distances from each other, and their relative $P$-values.

As with the observed objects, the mean plane of all Plutinos (black $\bigstar$) does not appear individually distinct from the mean plane of the $g+$ subset (green {\tiny $\blacktriangleright$}) or the $g-$ subset (yellow $\blacktriangledown$), because it lies within the 95\% probability contours for both subsets.
The mean planes of the librating subsets appear distinct from each other at their 95\% levels, but not at their 99.7\% levels as with the observed objects.
The mean plane of the $g+$ subset is distinct from the Neptune plane at about 95\% confidence and the mean plane of all Plutinos is distinct from the Neptune plane at high confidence, 
and the mean plane of the $g-$ subset is also distinct from the Neptune plane at high confidence.
All mean planes are widely separated from the Laplace plane, but none are distinct from the invariable plane.

\begin{centering}
\begin{figure*}
\begin{subfigure}{.33\textwidth}
  \centering
  \includegraphics[width=.8\linewidth]{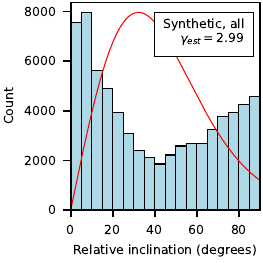}
  \label{fig:inclination_histograms_lawler_gplusminusnone}
\end{subfigure}%
\begin{subfigure}{.33\textwidth}
  \centering
  \includegraphics[width=.8\linewidth]{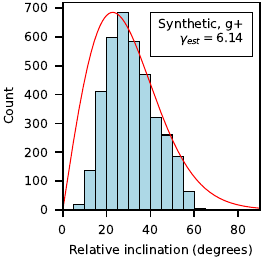}
  \label{fig:inclination_histograms_lawler_gplus}
\end{subfigure}%
\begin{subfigure}{.33\textwidth}
  \centering
  \includegraphics[width=.8\linewidth]{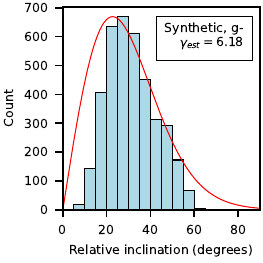}
  \label{fig:inclination_histograms_lawler_gminus}
\end{subfigure}%
\caption{Histograms of inclination for the simulated Plutinos relative to their vMF mean plane, and for their $g+$ and $g-$ librating subsets relative to their respective vMF mean planes.
The red curve is the vMF relative inclination distribution for the fitted concentration parameter.}
\label{fig:inclination_histograms_lawler}
\end{figure*}
\end{centering}

Based on the larger sample size of this synthetic sample compared to the current observational sample, we would expect the 95\% confidence region of the mean plane of all the simulated Plutinos to have an angular radius of about $2.04^\circ\times\sqrt{446/69,626}=0.17^\circ$, but instead the angular radius is $0.37^\circ$.
This may be due to the strong deviation of the inclinations from a vMF function and the high fraction of high-inclination objects in the simulated sample versus the small fraction of high-inclination objects in the observed population.
As shown in Figure \ref{fig:inclination_histograms_observed} and Figure \ref{fig:inclination_histograms_lawler}, the highest-inclination observed Plutino has $i=55.0^\circ$ relative to the \textsc{sct25} mean plane, whereas the fraction of simulated Plutinos with $i>55.0^\circ$ relative to their mean plane is $f=0.37$.
Similarly, the median inclinations of the $g$-librating subsets are slightly higher in the simulated population than in the observed population.
The greater inclination dispersion of the simulated population naturally leads to wider-than-expected mean plane confidence regions.

We note that the proportion of high-inclination objects in the population of all simulated Plutinos makes the vMF distribution used in the mean plane calculation a poor fit, as shown in Figure \ref{fig:inclination_histograms_lawler}: the red curve of the relative inclination distribution for the fitted vMF concentration parameter does not closely match the shape of the histogram.
However, the vMF confidence region for the mean plane still adequately describes the mean plane sampling distribution as long as the sample is sufficiently azimuthally symmetric around its mean plane; the simulated Plutinos satisfy this condition.
The relative inclination histograms of the simulated $g+$ and $g-$ librators more closely resemble the curves of the fitted vMF relative inclination distributions.
Though they have fewer low-inclination objects than the fitted distribution would suggest,
the match is close enough that we question whether the $g+$ and $g-$ librators of the simulated Lawler sample have the same relative inclination distributions as the $g+$ and $g-$ librators of the observed Plutino sample.
We run two-sample Anderson-Darling tests to compare the $g+$ relative inclinations for the observed and simulated Plutinos, and similarly for the $g-$ relative inclinations and for all relative inclinations.
In each case, the Anderson-Darling $P$-value is below 0.001, so we conclude that the uniformly-filled-parameter-space simulated sample does not match the observed Plutino population for the relative inclinations of the $g$ librators or for the Plutinos in general.

As with the observed Plutinos, the $g$-librating subsets of the simulated Plutinos have mean planes that are separated by $142^\circ$ of ascending node relative to the mean plane for all Plutinos and appear distinct from each other at the 95\% level, though neither appears individually statistically distinct from the mean plane of all simulated Plutinos.
The inclinations of the mean planes of the $g$-librating subsets, relative to the mean plane for all Plutinos (Table \ref{t:figure4_meanplanes}, second block, first row) are much smaller for the simulated objects than for the observed population, and the mean planes themselves are at much lower inclinations relative to the ecliptic plane and the invariable plane.
From Table \ref{t:figure4_meanplanes}, the mean planes of the $g+$ and $g-$ subsets of the synthetic Plutinos respectively have ecliptic-plane inclinations of $1.64^\circ$ and $2.41^\circ$, while they have invariable-plane inclinations of $0.60^\circ$ and $0.86^\circ$.
They are separated from the mean plane of all synthetic Plutinos by $0.63^\circ$ and $0.65^\circ$, respectively.
By contrast, from Table \ref{t:figure2_meanplanes}, the mean planes of the $g+$ and $g-$ subsets of the observed Plutinos respectively have ecliptic-plane inclinations of $6.93^\circ$ and $7.13^\circ$, with invariable-plane inclinations of $5.74^\circ$ and $7.17^\circ$ and separations from the mean plane of all observed Plutinos of $7.41^\circ$ and $5.47^\circ$.
The mean planes of all Plutinos have similar ecliptic inclinations ($1.88^\circ$ for the observed Plutinos and $1.82^\circ$ for the synthetic Plutinos), but different inclinations to the invariable plane ($1.77^\circ$ and $0.24^\circ$, respectively).

\end{document}